\documentclass[sn-mathphys,Numbered]{sn-jnl}

\usepackage{graphicx}%
\usepackage{multirow}%
\usepackage{amsmath,amssymb,amsfonts}%
\usepackage{amsthm}%
\usepackage{mathrsfs}%
\usepackage[title]{appendix}%
\usepackage{xcolor}%
\usepackage{textcomp}%
\usepackage{manyfoot}%
\usepackage{booktabs}%
\usepackage{algorithm}%
\usepackage{algorithmicx}%
\usepackage{algpseudocode}%
\usepackage{listings}%
\usepackage{hyperref}

\newcommand{\be}{\begin{equation}}
\newcommand{\ee}{\end{equation}}
\newcommand{\ba}{\begin{eqnarray}}
\newcommand{\ea}{\end{eqnarray}}
\newcommand{\rmi}[1]{{\mbox{\scriptsize #1}}}
\newcommand{\tab}{Tab.~}
\newcommand{\fig}{Fig.~}

\newcommand{\eq}{Eq.~}

\newcommand{\VT}{V^{ }_\rmi{$T$}}
\newcommand{\alphas}{\alpha_\rmi{s}}
\newcommand{\mD}{m_\rmi{D}}
\newcommand{\rmO}{{\mathcal{O}}}
\newcommand{\bmu}{\bar\mu}

\newcommand{\Nc}{N_{\rm c}}

\newcommand{\CF}{C_\rmi{F}}

\newcommand{\tinymsbar}{{\overline{\mbox{\tiny\rm{MS}}}}}
\newcommand{\Lambdamsbar}{{\Lambda_\tinymsbar}}

\newcommand{\arxiv}[2]{[arXiv:\,\href{http://arxiv.org/abs/#1}
{\texttt{#1}}[\texttt{#2}]]}

\theoremstyle{thmstyleone}%
\theoremstyle{thmstyletwo}%

\theoremstyle{thmstylethree}%

\begin{document}

\title[Article Title]{Study of quarkonium in QGP from unquenched lattice QCD}


\author*[1,2]{\fnm{Sajid} \sur{Ali}}\email{sajid.ali@physik.uni-bielefeld.de}

\author[1]{\fnm{Dibyendu} \sur{Bala}}\email{dibyendu.bala@physik.uni-bielefeld.de}
\equalcont{These authors contributed equally to this work.}

\author[1]{\fnm{Olaf} \sur{Kaczmarek}}\email{okacz@physik.uni-bielefeld.de}
\equalcont{These authors contributed equally to this work.}

\author[3]{\fnm{Hai-Tao} \sur{Shu}}\email{hai-tao.shu@ur.de}
\equalcont{These authors contributed equally to this work.}

\author[1]{\fnm{Tristan} \sur{Ueding}}\email{tueding@physik.uni-bielefeld.de}
\equalcont{These authors contributed equally to this work.}
\author[ ]{\\[1.5ex] (HotQCD collaboration)}

\affil*[1]{\orgdiv{Faculty of Physics}, \orgname{University of Bielefeld}, \orgaddress{\street{Universit\"atsstr.~25}, \city{Bielefeld}, \postcode{D-33615}, \state{NRW}, \country{Germany}}}

\affil[2]{\orgdiv{Department of Physics}, \orgname{Government College University}, \orgaddress{\street{Katchery Road}, \city{Lahore}, \postcode{54000}, \state{Punjab}, \country{Pakistan}}}

\affil[3]{\orgdiv{Institute for Theoretical Physics}, \orgname{University of Regensburg}, \orgaddress{\street{Universit\"atsstr.~31}, \city{Regensburg}, \postcode{D-93040}, \state{Bavaria}, \country{Germany}}}


\abstract{This paper discusses the charmonium and bottomonium correlators in the pseudoscalar channel and the corresponding spectral reconstruction on the lattice. The absence of a transport peak in the pseudoscalar channel spectral function allows for an easier study of the in-medium modification of bound states. However, extracting spectral information from Euclidean correlators is still a numerically ill-posed problem. To constrain the spectral reconstruction, we use an ansatz motivated from perturbation theory. The perturbative model spectral function has two main contributions: a thermal part around the threshold obtained from pNRQCD and the vacuum part well above the threshold. These two regions are matched continuously, and the model spectral function is obtained by introducing parameters that control the overall thermal shift of the peak and the overall amplitude. The lattice correlator data is computed using clover-improved Wilson valence fermions on large and fine gauge field configurations generated using $N_f=2+1$ flavors Highly Improved Staggered Quark (HISQ) action with physical strange quark mass $m_s$, and slightly heavy degenerate up and down quark masses $m_l=m_s/5$ that correspond to $m_\pi\simeq 320$ MeV. Our results obtained at $T=220$ MeV and $T=251$ MeV suggest that no resonance peaks are needed to describe the charmonium lattice data at these temperatures, while for bottomonium thermally broadened resonance peaks persist.}




\maketitle

\section{Introduction}\label{sec:intro}

Heavy quark-antiquark bound states, quarkonia, like charmonium and bottomonium, serve as a good thermometer for the quark-gluon-plasma (QGP) created in ultra-relativistic heavy ion (AA) collisions~\cite{Mocsy:2008eg}. This is due to the fact that they are produced during the early stages of the collisions and participate the entire evolution of the QGP. Some of them will dissociate into heavy quarks when experiencing temperature increasing while some can remain as bound states due to their hierarchically small sizes and large binding energies~\cite{Matsui:1986dk,Karsch:2005nk}. How to theoretically understand the suppression of quarkonia yields observed in AA collisions compared to that in the proton-proton (pp) collisions~\cite{CMS:2017ycw,CMS:2018zza,ALICE:2018wzm}, however, remains unclear. This is mainly due to the complicated entanglement between the cold and hot nuclear effects~\cite{Brambilla:2010cs}. Given that the suppression resides in the non-perturbative regime, naturally we want to interpret it using lattice QCD calculations. Besides, the experiments at RHIC and LHC reveal that open heavy mesons show an unexpectedly substantial elliptic flow, comparable to that of light mesons~\cite{PHENIX:2006iih,STAR:2006btx,ALICE:2012ab}. This indicates that heavy quarks flow as efficiently as light quarks do~\cite{ALICE:2013olq,PHENIX:2014rwj,Vertesi:2014tfa}. Trying to interpret this requires a modeling of the heavy quark diffusion in a hot and dense medium, which is challenging in perturbation theory~\cite{He:2014epa,Cao:2014fna,Cao:2018ews,Li:2021xbd}, but feasible from lattice QCD. Studies carried out in the heavy quark mass limit can be found in \cite{Francis:2015daa,Brambilla:2020siz,Altenkort:2020fgs,Banerjee:2022gen,Brambilla:2022xbd,Altenkort:2023oms} in both quenched approximation and (2+1)-flavor lattice QCD. Accessing the first mass suppressed correction to the diffusion coefficient are also attempted~\cite{Bouttefeux:2020ycy,Banerjee:2022uge,Brambilla:2022xbd,Altenkort:2021ntw}.

Tackling the above puzzles relies on the knowledge about the spectral function of quarkonia, which can be extracted from the fully relativistic Euclidean mesonic correlation functions. The spectral function in the vector channel contains all information about the in-medium hadron properties like the dissociation temperatures and heavy quark diffusion, and was intensely studied in the past decade, see~\cite{Ding:2012sp,Ding:2017std,Kim:2018yhk, Ding:2021ise,Aarts:2011bottom,Aarts:2013pwave,Aarts:2014bspectrum} for a selection. While the one in the pseudoscalar channel provides an easier access to the fate of the bound states due to the vanishing of the transport peak \cite{Karsch:2003wy,Aarts:2005hg}. This was investigated in the quenched case in \cite{Burnier:2017bod}. A review of current status of lattice studies on heavy quarks and quarkonium in extreme conditions can be found in~\cite{Rothkopf:2019ipj,Ding:2020rtq,Kaczmarek:2022ffn}.

In this work we aim to provide a (2+1)-flavor lattice QCD calculation for the charmonium and bottomonium correlation functions in pseudoscalar channel, $G_\rmi{PS}(\tau)$, at two temperatures $T$=220 MeV and $T$=251 MeV. 
The Euclidean correlator is related to the spectral function via following equation
\be
  G^{ }_\rmi{PS}(\tau) \; = \; 
 \int_0^\infty
 \frac{{\rm d}\omega}{\pi} \rho^{ }_\rmi{PS} (\omega)
 \frac{\cosh \left( \left(\frac{1}{2T} - \tau\right)\omega \right) }
 {\sinh\left( \frac{\omega}{2 T} \right) } 
 \;.\label{eq-corr-spec}
\ee
Due to statistical errors in $G_\rmi{PS}(\tau)$ and a finite resolution of the lattice, the solution of $\rho^{ }_\rmi{PS}$ is not unique. However, the spectral reconstruction can be constrained with further inputs. In this study we model the spectral function with a perturbatively inspired ansatz, interpolating between the thermal part around the threshold obtained from pNRQCD~\cite{Laine:2007gj} and the vacuum part well above the threshold~\cite{Burnier:2017bod}.

This article is organized as follows. We begin with the lattice setup and clover mass tuning, which are outlined in Sec.~\eqref{sec:param}. Sec.~\eqref{sec:corr} presents the numerical data of lattice correlators, while Sec.~\eqref{sec:pert} details the process of obtaining perturbative spectral functions. In addition, Sec.~\eqref{sec:recon} provides an explanation of the spectral reconstruction procedure. Finally, we summarize our findings and present future outlook in the concluding section.
\section{Lattice parameters and mass tuning}\label{sec:param}

Quarkonium correlators are obtained at various temperatures, as specified in \tab\ref{tab-latticesetup}, using clover-improved Wilson fermions as valence quarks. 
The lattice correlation functions are computed on large and fine gauge field configurations generated with $N_f=2+1$ and with the 
Highly Improved Staggered Quark (HISQ) action~\cite{Follana:2006rc} and tree-level improved L\"uscher-Weisz gauge action~\cite{Luscher:1984xn,Luscher:1985zq}. The sea quark masses are tuned to the physical strange quark mass, $m_s$, and degenerate up, down quark masses $m_l=m_s/5$, corresponding to $m_\pi\simeq 320$ MeV.
The fermion valence action is tadpole-improved by incorporating clover-improvement, which reduces the $\mathcal{O}(a)$ cut-off effects. To this aim, the Sheikholeslami-Wohlert coefficient~\cite{Sheikholeslami:1985ij} is chosen as $c_{SW}=1/u^3_0$, where $u_0$ is the tadpole factor determined by taking the fourth root of the plaquette expectation value calculated on our HISQ lattices.\par
\begin{figure}
\centering
\includegraphics[width=0.6\textwidth]{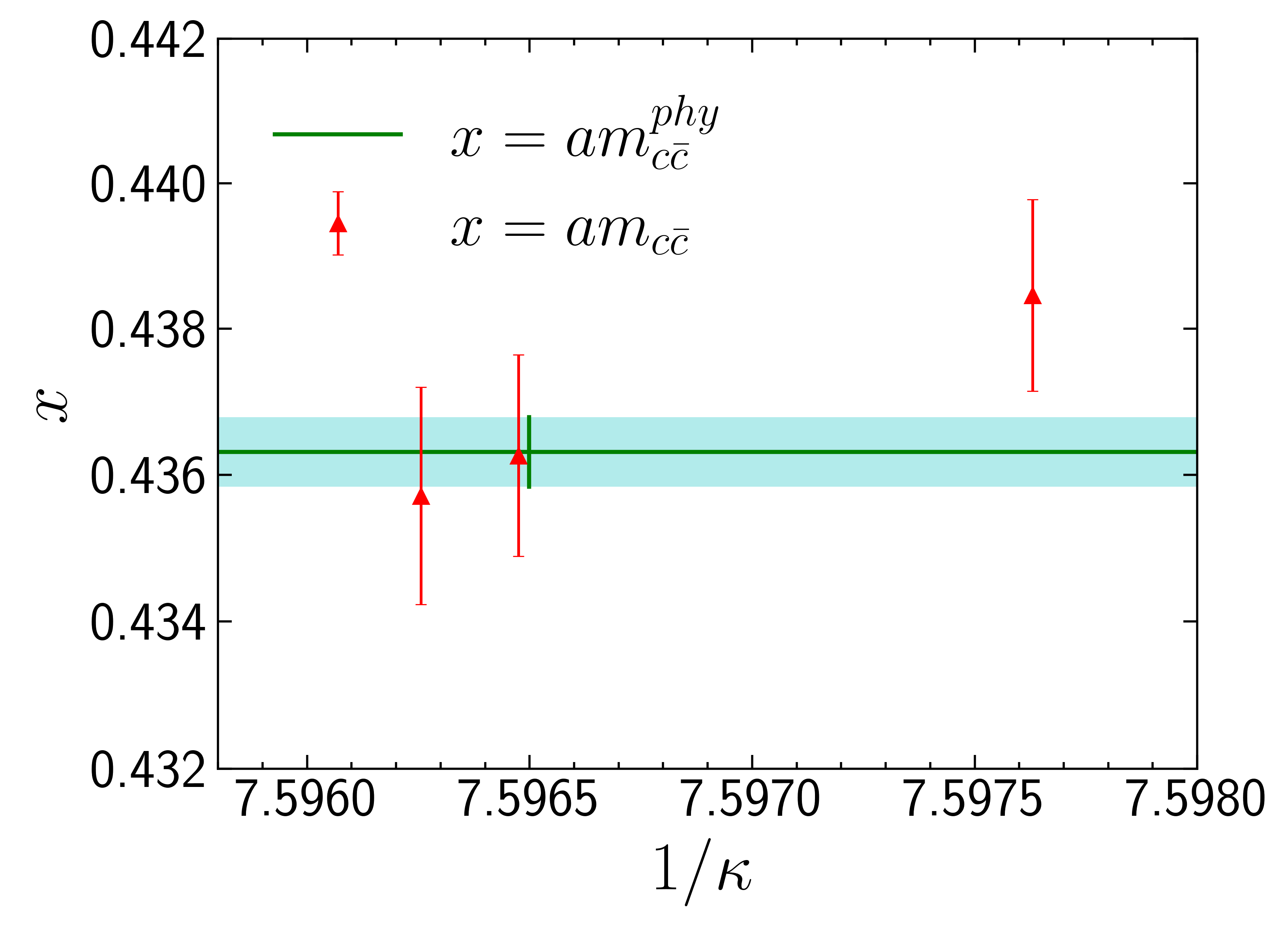}
\caption{Charm quark mass tuning on the mixed action. The solid line represents the 
experimental value of the spin-averaged charmonium in lattice units taken from the PDG~\cite{ParticleDataGroup:2022pth}. 
Whereas triangles represent masses from lattice calculations for different values of 
$\kappa$ in lattice units.}
\label{tuning}
\end{figure}

\begin{table}
\begin{tabular}{@{}lllllll@{}}
\toprule
$\beta$ & $a$[fm] & $a^{-1}$[GeV] & $N_\sigma$ & $N_\tau$ & $T$[MeV] & $\#$ confs\\
\midrule
\multirow{3}{*}{8.249} & \multirow{3}{*}{0.028} & \multirow{3}{*}{7.033} & 64 & 64 & 110 & 112\\
&& & 96 & 32 & 220 & 1703\\  
&& & 96 & 28 & 251 & 621\\
\botrule 
\end{tabular}
\vspace{1mm}
\caption{Lattice parameters for $N_f$=2+1, $m_l=m_s/5$ HISQ configurations for different temperatures. The lattice spacing $a$ at $\beta$=8.249 is obtained using $f_K$-scale parametrization, cf. Ref.~\cite{Bazavov:2014eq}. The tuning of the quark masses is done at lowest temperature with maximum temporal extent $N_\tau$=64.} 
\label{tab-latticesetup}
\end{table}
The masses of the lattice calculated correlators are determined by the hopping parameter $\kappa$, which should be tuned to match the quarkonium meson mass spectrum observed in experiments~\cite{ParticleDataGroup:2022pth}. In the heavy quark sector, the tuning is performed with respect to the spin-averaged charmonium $m_{c\bar{c}}=(m_{\eta_{c}}+3m_{J/\psi})/4$ and bottomonium $m_{b\bar{b}}=(m_{\eta_{b}}+3m_{\Upsilon})/4$. \fig\ref{tuning} illustrates the charm quark mass tuning, where $am^{phy}_{c\bar{c}}$ represents the physical mass in lattice units. The values of $\kappa$ that are tuned for charm and bottom are 0.13164 and 0.11684, respectively.
\section{Correlators}\label{sec:corr}

\begin{figure}[hbt!]
\centering
\includegraphics[width=0.49\textwidth]{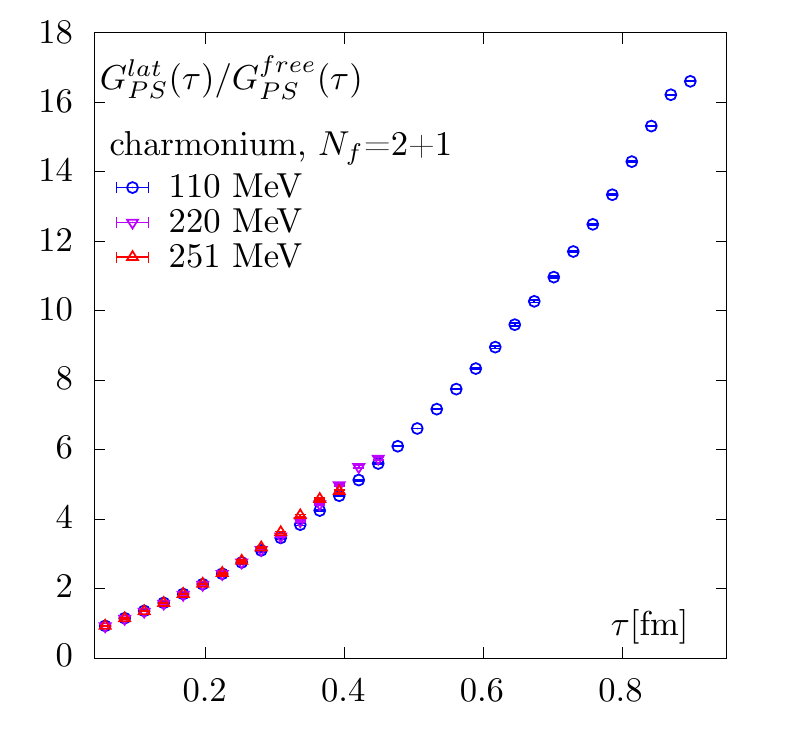}
\includegraphics[width=0.49\textwidth]{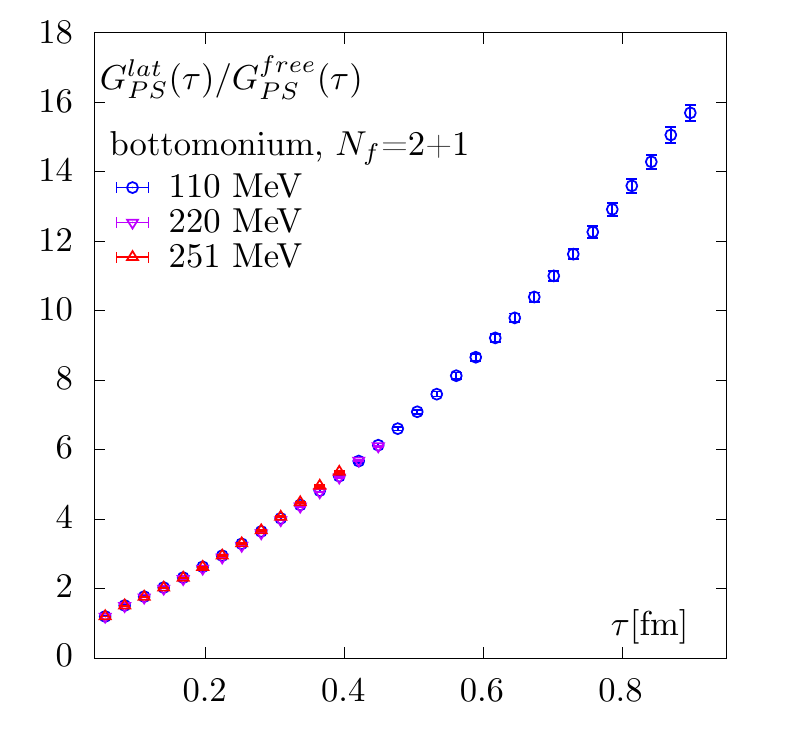}
\caption{The ratio of pseudoscalar lattice correlator with its corresponding free correlator at different temperatures for charmonium (left) and bottomonium (right). Statistical errors are estimated using the Jackknife procedure.}
\label{fig-corr}
\end{figure}

We present the pseudoscalar channel charmonium and bottomonium  correlators in Fig. \ref{fig-corr}. Instead of the correlator itself, we used the ratio 
$G_\rmi{PS}(\tau)/G^\rmi{free}_\rmi{PS}(\tau)$ for a better visibility of the behavior as a 
function of $\tau$ at various temperatures, where $G^\rmi{free}_\rmi{PS}(\tau)$ reads \cite{Burnier:2017bod} 
\ba
 \frac{ G_\rmi{PS}^\rmi{free}(\tau) }{ m^2(\bmu^{ }_\rmi{ref}) } 
 & \equiv &   
 \int_{2 M_q}^{\infty}
 \! \frac{{\rm d}\omega}{\pi} \, 
 \biggl\{ 
 \frac{\Nc \omega^2}{8\pi}
 \tanh\Bigl( \frac{\omega}{4T} \Bigr)  
 \sqrt{1 - \frac{4 M_q^2}{\omega^2}} 
 \biggr\}
 \, 
 \frac{\cosh \left(\left(\frac{1}{2T} - \tau\right)\omega\right)}
 {\sinh\left(\frac{\omega}{2T}\right)} 
 \;, \label{GPSfree}
\ea
where $m(\bar{\mu}_\rmi{ref})$ denotes the running mass at reference scale 
$\bar{\mu}_\rmi{ref}$=2 GeV. We choose the values of $M_q$ to be 1.5 GeV for charmonium and 
4.7 GeV for bottomonium. One can see that the charmonium 
suffers more thermal effects than the bottomonium does, because the charmonium correlators at high temperatures deviate more from the low-temperature ones. 
%
\section{Perturbative spectral functions in PS-channel}\label{sec:pert}

To construct the quarkonium spectral functions in the pseudoscalar channel we follow the strategy developed in the quenched approximation~\cite{Burnier:2017bod}. For $\omega \gg 2 M$, the spectral function is insensitive to temperature effects, therefore it is replaced by its vacuum counterpart. The vacuum spectral function, normalised by $\omega^2 m^2(\bar{\mu})$, in the pseudoscalar channel reads
\be
 \left. \frac{ \rho^{ }_\rmi{PS}(\omega) }{\omega^2 m^2(\bar{\mu})} \right|^\rmi{vac} 
 \; \equiv \; 
 \frac{\Nc}{8\pi}\, \tilde{R}_\rmi{c}^{p}(\omega, \bar \mu) 
 \;.\label{def_v}
\ee
Here $\tilde{R}$-function is valid up to $\rmO(\alphas^3)$ and it is defined in Ref.~\cite{Burnier:2017bod}. One can also express the spectral function $\rho^{}_\rmi{PS}$ normalized by the pole mass $M$ as follows:
\be
 \left. \frac{ \rho^{ }_\rmi{PS}(\omega) }{\omega^2 M^2} \right|^\rmi{vac} 
 \; \equiv \; 
 \frac{\Nc}{8\pi}\, {R}_\rmi{c}^{p}(\omega) 
 \;.\label{def_v_}
\ee
where $R$ is again given in Ref.~\cite{Burnier:2017bod}. However, this expansion does not converge as the correction term becomes increasingly large as $\omega$ increases. In contrast, the form in Eq.~\eqref{def_v}, obtained by re-expanding the pole mass in terms of the $\overline{MS}$ mass in Eq.~\eqref{def_v_}, leads to  smaller corrections at large $\omega$ . 
Following Ref.~\cite{Burnier:2017bod}  we can then estimate the pole mass self-consistently by equating Eq.~\eqref{def_v}  and Eq.~\eqref{def_v_}, which is given by,
\be
 M_x^2 \equiv m^2(\bmu)\,
 \left.  \frac{\tilde{R}^p_\rmi{c}(\omega,\bmu)}{R^p_\rmi{c}(\omega)}
 \right|_{\bmu=\omega,~ \omega = x M^{ }_x}
 \;. \label{fixmass}
\ee
Here the running of the $\overline{MS}$ mass has been calculated using 5-loop running, starting from $m(\bar{\mu}= 2 \, \text{GeV})$. \tab\ref{table:masses} shows the values of $M_x$ obtained at $\rmO(\alphas^2)$.\par
\begin{table}[h]
\begin{tabular}{@{}llll@{}}
 \toprule
 $\displaystyle
  \frac{m(\bmu^{ }_\rmi{ref} \equiv \mbox{2~GeV})}{\rmi{GeV}}$ &
 $\displaystyle
  \frac{m(\bmu = m)}{\rmi{GeV}}$ &
 $\displaystyle
  \frac{M^{ }_x}{\rmi{GeV}}$ & 
 $\displaystyle
  \alphas(m(\bmu = m))$ \\
 \midrule
  1.08 & 1.28 & 1.35(1) & 0.400 \\
  5.00 & 4.21 & 4.50(2) & 0.217 \\
 \bottomrule
\end{tabular}
\vspace{2mm}
\caption{The quark masses relevant for the perturbative spectral functions for full QCD, utilizing 5-loop running for $\alphas$ and setting $\Lambdamsbar$ to 0.339 GeV~\cite{Aoki:202flag}. The mass $M_x$ was defined using Eq.~\eqref{fixmass}, with $x = 4...8$. For the charmonium case, we selected $m(\bar{\mu}_\rmi{ref})$ = 1.08 GeV to ensure that $m(\bmu = m)$ yields mass in the $\overline{\text{MS}}$ scheme that is relevant. On the other hand, a value of $m(\bmu^{ }_\rmi{ref})$ = 5 GeV is close to the value relevant for the bottomonium case.}
\label{table:masses}
\end{table}
Now we consider thermal part of the spectral function around the threshold i.e. $\omega \approx 2 M$. In the heavy quark limit, relativistic effects in quark-antiquark bound states become negligible, allowing for a non-relativistic potential description. For separations much smaller than $1/gT$, the potential is effectively a zero temperature potential. However, as the separation approaches $\sim 1/gT$, the contribution of soft gluon exchange becomes significant, requiring a Hard Thermal Loop resummation. As a result, the potential in this region becomes temperature dependent and acquires a complex component. In this non-relativistic regime, one can get the pseudoscalar spectral function from the vector channel as 
\be
 \rho^\rmi{pNRQCD}_\rmi{PS} = 
 \frac{M^2}{3}
 \rho^\rmi{pNRQCD}_\rmi{V}
 \;.  \label{factor2}
\ee
The vector channel spectral function is given by
\be
 \rho^\rmi{pNRQCD}_\rmi{V}(\omega) = 
 \frac12 \Bigl( 1 - e^{-\frac{\omega}{T }}\Bigr)
 \int_{-\infty}^{\infty} \! {\rm d} t \, e^{i \omega t}
 \; C_{>}(t;{\vec{0},\vec{0}})
 \;,  \label{nrqcd_T}
\ee
where $C^{ }_{>}$ is a Wightman function and it is obtained by solving the following Schr\"odinger equation 
\be
 \biggl\{ i \partial_t - \biggl[ 2 M 
 + \VT(r) - \frac{\nabla^2_{\vec{r}}}{M}
 \biggr] \biggr\} \, C_{>}^V(t;{\vec{r},\vec{r'}}) = 0 
 \;,  \quad t\neq 0 \;, \label{Seq}
 \ee
 with initial conditions 
 $C_{>}^V(0;{\vec{r},\vec{r'}}) = 6 \Nc\, \delta^{(3)}({\vec{r}-\vec{r'}})\;$. The potential $\VT(r)$ for positive $t$ has the following form~\cite{imV,bbr,jacopo}
\be
  \VT(r) = -\alphas C_F \biggl[ \mD^{ } + \frac{\exp(-\mD^{ } r)}{r}
 \biggr] - {i \alphas \CF T } \, \phi(\mD^{ } r)  + \rmO(\alphas^2)\;. \label{expl}
\ee
Here $\phi(x)$ is defined to be
\be
 \phi(x) \equiv 
 2 \int_0^\infty \! \frac{{\rm d} z \, z}{(z^2 +1)^2}
 \biggl[
   1 - \frac{\sin(z x)}{zx} 
 \biggr].
 \label{phi}
\ee
Here $\alpha_\rmi{S}$ is the thermal coupling and $m_{\rmi{D}}$ $(\sim gT)$ is the Debye mass.
Physically, the real part of the potential is related to Debye screening in the plasma. It affects the thermal mass shift of bound states. On the other hand, the imaginary part originates from Landau damping of space like gauge fields, which causes the bound state peak to broaden.
At a much lower $\omega$ than the threshold, the Schr\"odinger description described above overestimates the spectral function, as this formalism is no longer valid in this region. A naive NLO calculation shown in Ref.~\cite{Burnier:2017bod} predicts that the spectral function in this region is exponentially suppressed. To model this suppression within the Schr\"odinger formalism itself, we multiply the imaginary part of the potential by $e^{-|\omega-2\,M|/T}$ for $\omega<2\,M$. Moreover, at small spatial distances, $r \ll 1/\mD$, the thermal potential is replaced by a vacuum one \cite{Burnier:2017bod}.\par
\begin{figure}[hbt!]
\centering
\includegraphics[width=0.49\textwidth]{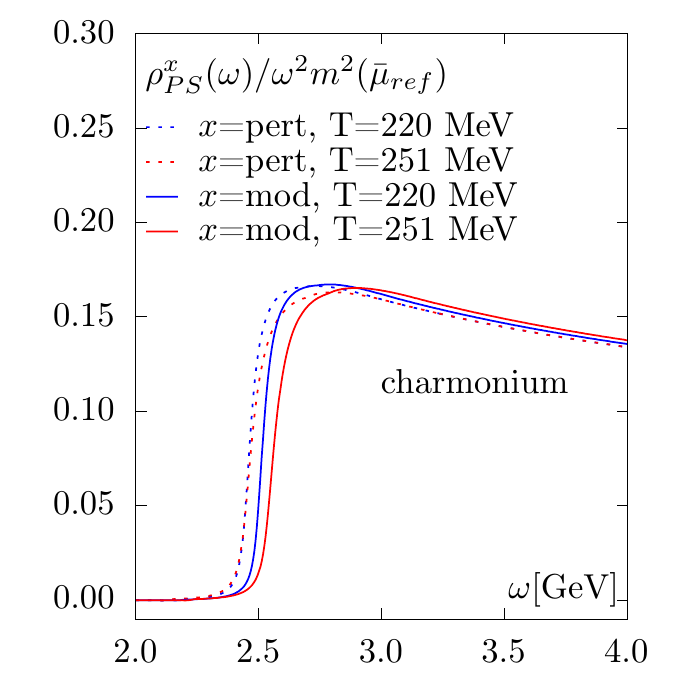}
\includegraphics[width=0.49\textwidth]{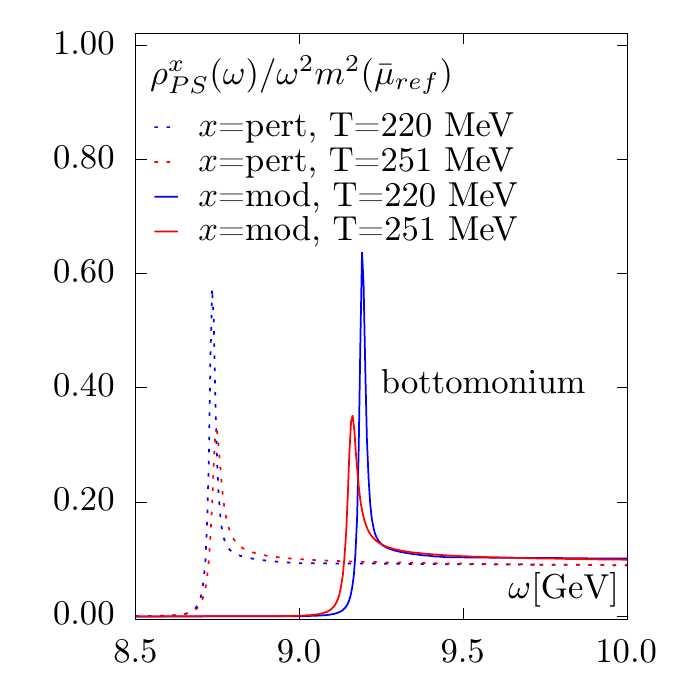}
\caption{
The comparison between perturbative and model spectral functions. The dotted lines represent the perturbative spectral functions, while the solid lines illustrate their modifications as per Eq.~\eqref{eq:spfmod}, for both charmonium (on the left) and bottomonium (on the right). The solid curves correspond to the imaginary-time correlators, labeled as ``$x$=mod" in Fig.~\ref{fig:modlatcorr}, which exhibit good agreement with the lattice data for all distances, except at small distances where the lattice cut-off effects dominate due to the finite lattice spacing.}\label{fig:purtsp}
\end{figure}
After computing both thermal and vacuum contributions to the spectral function, we
combine them as
\begin{align}
\label{eq-pertspf}
\begin{split}
\rho_\rmi{PS}^\rmi{pert}(\omega)= A^\rmi{match}\rho^{\rmi{pNRQCD}}_\rmi{PS}(\omega)\theta(\omega^\rmi{match}-\omega)
+\rho^\rmi{vac}_\rmi{PS}(\omega)\theta(\omega-\omega^\rmi{match})\;.
\end{split}
\end{align}
In \eq(\ref{eq-pertspf}) we introduce a multiplicative factor $A^\rmi{match}$ to the 
thermal part and it is determined such that both thermal and vacuum parts of the 
spectral function are connected smoothly at some $\omega=\omega^\rmi{match}$, where $2M<\omega^\rmi{match}<3M$. 
The resulting pseudoscalar spectral functions at various temperatures for both
charmonium and bottomonium are shown in Fig. \ref{fig:purtsp}. Unlike charmonium, the bottomonium has a resonance peak that is visible at temperatures 220 MeV and 251 MeV. The resonance peak for bottomonium hints the existence of bound state at these given temperatures, however the peak might get broaden at higher temperatures.
\section{Modelling and spectral reconstruction}\label{sec:recon}

In this section, we discuss the process of modeling perturbative spectral functions. We introduce two parameters, $A$ and $B$, in the perturbative spectral function shown in Fig. \ref{fig:purtsp}. Parameter $A$ will account for the normalization of the correlator data, while parameter $B$ will allow for the adjustment of thermal mass shift because pole masses are poorly determined in perturbation theory. As a results the model spectral function gets the following form:
\be
\rho^\rmi{mod}_\rmi{PS}(\omega)=A\rho^\rmi{pert}_\rmi{PS}(\omega-B)\;.
\label{eq:spfmod}
\ee
Before fitting the lattice data with the above ansatz, it is important to note that the pseudoscalar correlator can be multiplicatively renormalized, which only affects the fit parameter $A$. Since our analysis involves correlators at different temperatures obtained from a single lattice spacing, the renormalization constant will be the same for all temperatures. Therefore, we do not need to renormalize the lattice correlator for our present study. Nevertheless, to avoid obtaining large values of $A$ after fitting, we normalize the lattice correlator such that it matches the perturbative correlator at a very short distance of $\tau T = 0.107$ at a temperature $T=220$ MeV.

\begin{figure}[hbt!]
\centering
\includegraphics[width=0.49\textwidth]{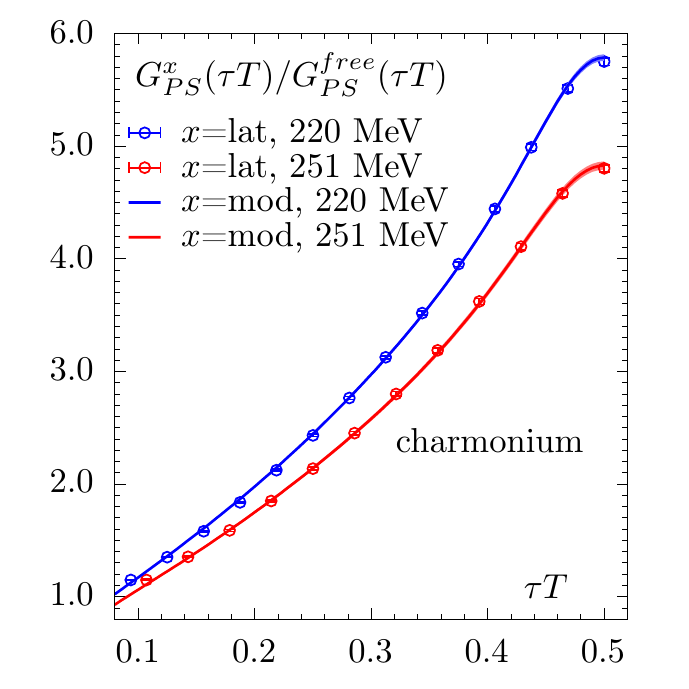}
\includegraphics[width=0.49\textwidth]{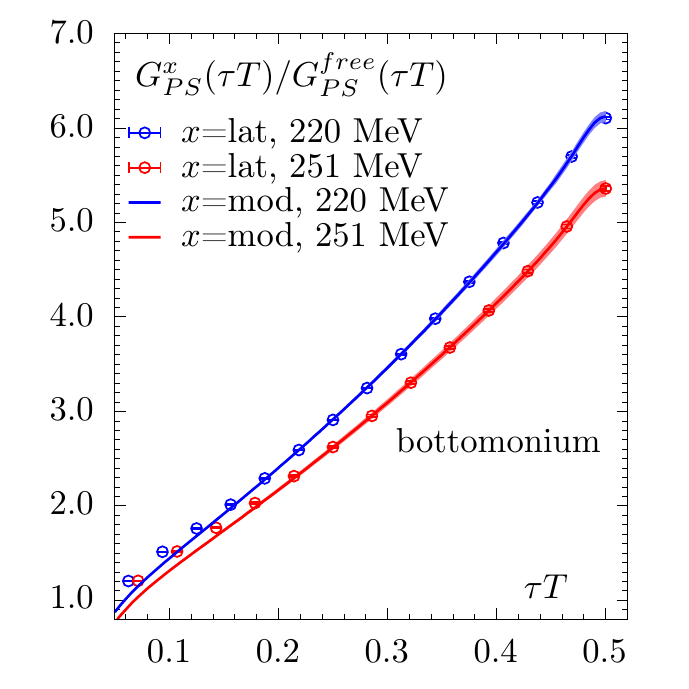}
\caption{The data points correspond to the lattice correlators, while the lines depict the model correlators obtained from \eq(\ref{eq-corr-spec}) for charmonium (on the left) and bottomonium (on the right).}
\label{fig:modlatcorr}
\end{figure}

We perform uncorrelated $\chi^2$ fit of this normalized lattice data with the ansatz in Eq.~\eqref{eq:spfmod}. The short distance part of the lattice correlator are largely affected by lattice artifacts. As a result, we fit the data starting from $\tau T$=0.25 for $T=220$ MeV and $\tau T$=0.21 for $T=251$ MeV. The resulting fit parameters, $A$ and $B$, for two distinct temperatures, are presented in~\tab\ref{tab:fitparam}. Preliminary results of the model spectral function $\rho^\rmi{mod}$ and the lattice correlator data, are shown in Fig. \ref{fig:modlatcorr}.
\begin{table}[t]
\begin{tabular*}{\textwidth}{@{\extracolsep\fill}lcccccc}
 \toprule 
 & \multicolumn{3}{@{}c@{}}{charmonium} 
 & \multicolumn{3}{@{}c@{}}{bottomonium} \\\cmidrule{2-4}\cmidrule{5-7}
 $\displaystyle {T}$  (MeV) &
 $\displaystyle A $ &
 $\displaystyle { B } / { T } $ & 
 $\displaystyle {\chi^2} / {\mbox{d.o.f.}} $ &
 $\displaystyle A $ &
 $\displaystyle { B } / { T } $ &  
 $\displaystyle {\chi^2} / {\mbox{d.o.f.}} $\\
 \midrule
  $220$ & 1.0139(42) & 0.368(11) & 0.89  & 1.1115(37) & 2.0824(91) & 0.86  \\ 
  $251$ & 1.005(11)  & 0.239(19) & 0.28  & 1.0877(78) & 1.667(13)  & 0.05  \\ 
  \bottomrule
\end{tabular*}
\vspace{2mm}
\caption{Estimates of the best fit parameters at two temperatures according to Eq.~\eqref{eq:spfmod}
  for charmonium and bottomonium.}
\label{tab:fitparam}
\end{table}
The important observation is the shift of the thresholds towards larger energy in the model spectral function. Overall, the observed shift is perfectly admissible and could be due to the uncertainties in the determination of the pole mass or the thermal mass corrections. It is important to take these uncertainties into account when interpreting the results.
\section{Conclusion and Outlook}

In this work we have presented preliminary results of the spectral reconstruction in the pseudoscalar channel for quarkonium, using an ansatz for the fitting procedure based on perturbation theory. We solved the Schr\"odinger equation for the thermal part of the spectral function around the threshold, using the perturbative finite temperature potential. For energies well above the threshold, we considered the vacuum part and matched both parts smoothly. Our analysis was limited to two temperatures, namely $T=220$ MeV and $T=251$ MeV, and we obtained lattice correlators using clover-improved Wilson fermions measured on large $N_f=2+1$ HISQ gauge field configurations generated at gauge coupling $\beta=8.249$. we have tuned the quark masses so that the experimental meson spectrum was reproduced on the lattice within errors.

Fitting the spectral function model to the lattice data, we observe a good description of the lattice correlators suggesting that no resonance peaks are needed to describe the charmonium lattice data at these temperatures, while for bottomonium thermally broadened resonance peaks persist at the analyzed temperatures.
While our model describes the lattice data well for large time-slice separations, we observed a discrepancy at small distances, which probably is due to the cut-off effects. To estimate these effects, we are in the process of adding more lattice spacings followed by a continuum extrapolation. The non-perturbative effects are important in the temperature range we are considering in this work. Therefore, to improve the model of the spectral function for this range in the future, we will estimate the thermal part of the spectral function using non-perturbatively determined thermal potential from the lattice \cite{BKR,BD,BKL}. Additionally, we plan to estimate the pole mass, which affects the thermal part of the spectral function and is responsible for the peak position, from the zero-temperature Cornell potential~\cite{BKR1}. This will help us better understand the thermal mass shift.

Furthermore, we plan to include lattices at physical quark masses and to consider the vector channel of the quarkonium spectral function, from which we will be able to estimate the heavy quark diffusion coefficients.
\bmhead{Acknowledgments}

We would like to express our appreciation to Mikko Laine for the valuable discussions we had over email. Additionally, we extend our appreciation to Luis Altenkort for generating gauge field configurations and integrating meson measurements into the QUDA code. This work is supported by the Deutsche Forschungsgemeinschaft (DFG, German Research Foundation)-Project number 315477589-TRR 211. The computations in this work were performed on the GPU cluster at Bielefeld University using \texttt{SIMULATeQCD} suite \cite{Bollweg:2021cvl,Mazur:2021zgi} and QUDA~\cite{Clark:2009wm}. We thank the Bielefeld HPC.NRW team for their support.



\end{document}